\documentclass[prb,letterpaper,twocolumn,showpacs,superscriptaddress,amsmath,amsfonts,amssymb,preprintnumbers,citeautoscript,nobalancelastpage]{revtex4}
\pdfoutput=1
\usepackage[T1]{fontenc}\usepackage[latin1]{inputenc}
\usepackage{dcolumn,graphicx,color,booktabs,microtype,afterpage} 
\usepackage[charter]{mathdesign}
\renewcommand{\figurename}{Fig.}
\renewcommand{\tablename}{Table}
\makeatletter\renewcommand{\fnum@figure}[1]{\figurename~\thefigure~(color online). }\makeatother
\makeatletter\renewcommand{\fnum@table}[1]{\tablename~\thetable. }\makeatother

\usepackage[colorlinks,plainpages=false,linkcolor=blue,urlcolor=blue,citecolor=black,pdfpagemode=UseNone,pdfstartview=FitBH]{hyperref}

\begin{document}\pagestyle{plain}

\title{Symmetry and disorder of the vitreous vortex lattice in an overdoped BaFe$_{2-x}$Co$\kern-1pt_x\kern1pt$As$_2$ superconductor: Indication for strong single-vortex pinning}

\author{D.\,S.\,Inosov}\email[Corresponding author: \vspace{4pt}]{d.inosov@fkf.mpg.de}
\affiliation{Max Planck Institute for Solid State Research, Heisenbergstra{\ss}e~1, 70569 Stuttgart, Germany}

\author{T.~Shapoval}\author{V.~Neu}\author{U.\,Wolff}
\affiliation{Leibniz Institute for Metallic Materials, IFW~Dresden, Helmholzstra{\ss}e 20, 01069 Dresden, Germany}

\author{J.\,S.\,White}
\affiliation{School of Physics and Astronomy, University of Birmingham, Edgbaston, Birmingham, B15~2TT, UK}

\author{S.\,Haindl}
\affiliation{Leibniz Institute for Metallic Materials, IFW~Dresden, Helmholzstra{\ss}e 20, 01069 Dresden, Germany}

\author{J.\,T.~Park}\author{D.\,L.\,Sun}\author{C.\,T.~Lin}
\affiliation{Max Planck Institute for Solid State Research, Heisenbergstra{\ss}e~1, 70569 Stuttgart, Germany}

\author{E.\,M.\,Forgan}
\affiliation{School of Physics and Astronomy, University of Birmingham, Edgbaston, Birmingham, B15~2TT, UK}

\author{M.\,S.\,Viazovska}
\affiliation{Max Planck Institute for Mathematics, Vivatsgasse 7, 53111 Bonn, Germany}

\author{J.\,H.\,Kim}
\affiliation{Max Planck Institute for Solid State Research, Heisenbergstra{\ss}e~1, 70569 Stuttgart, Germany}

\author{M.\,Laver}
\affiliation{Laboratory for Neutron Scattering, Paul Scherrer Institut, 5232 Villigen PSI, Switzerland}
\affiliation{University of Maryland, College Park, Maryland, 20742 Maryland, USA}

\author{K.\,Nenkov}
\affiliation{Leibniz Institute for Metallic Materials, IFW~Dresden, Helmholzstra{\ss}e 20, 01069 Dresden, Germany}

\author{O.\,Khvostikova}
\affiliation{Leibniz Institute for Complex Materials, IFW~Dresden, Helmholzstra{\ss}e 20, 01069 Dresden, Germany}

\author{S.\,Kühnemann}
\affiliation{Max Planck Institute for Metals Research, Stuttgart Center for Electron Microscopy, Heisenbergstra{\ss}e~3, 70569 Stuttgart, Germany}

\author{V.~Hinkov}
\affiliation{Max Planck Institute for Solid State Research, Heisenbergstra{\ss}e~1, 70569 Stuttgart, Germany}

\begin{abstract}
\noindent The disordered flux line lattice in single crystals of the slightly overdoped BaFe$_{2-x}$Co$_x$As$_2$ ($x=0.19$, $T_{\rm c}=23$\,K) superconductor is studied by magnetization measurements, small-angle neutron scattering (SANS), and magnetic force microscopy (MFM). In the whole range of magnetic fields up to 9\,T, vortex pinning precludes the formation of an ordered Abrikosov lattice. Instead, a vitreous vortex phase (vortex glass) with a short-range hexagonal order is observed. Statistical processing of MFM datasets lets us directly measure its radial and angular distribution functions and extract the radial correlation length $\zeta$. In contrast to predictions of the collective pinning model, no increase in the correlated volume with the applied field is observed. Instead, we find that $\zeta$ decreases as $(1.3\pm0.1)\,R_1\propto H^{-1/2}$ over four decades of the applied magnetic field, where $R_1$ is the radius of the first coordination shell of the vortex lattice. Such universal scaling of $\zeta$ implies that the vortex pinning in iron arsenides remains strong even in the absence of static magnetism. This result is consistent with all the real- and reciprocal-space vortex-lattice measurements in overdoped as-grown BaFe$_{2-x}$Co$_x$As$_2$ published to date and is thus sample-independent. The failure of the collective pinning model suggests that the vortices remain in the single-vortex pinning limit even in high magnetic fields up to 9\,T.
\end{abstract}

\keywords{iron arsenide superconductors, vortex pinning, magnetic force microscopy, small-angle neutron scattering}
\pacs{74.70.Xa 74.25.Wx 68.37.Rt 61.05.fg}

\maketitle

\vspace{-5pt}\section{Introduction}\vspace{-5pt}

The behavior of supercurrent vortices associated with magnetic flux quanta penetrating a \mbox{type-II} superconductor, known as Abrikosov vortices \cite{Abrikosov57}, contains valuable information about the microscopic material properties, such as the symmetry of the superconducting order parameter or the vortex pinning force \cite{BlatterFeigelman94, SonierBrewer00}. These, in their turn, determine essential macroscopic physical characteristics of the superconductor, such as the critical current density or critical fields. In clean, highly homogeneous superconductors the vortices form an ordered Abrikosov lattice due to the repulsive intervortex interactions, as was first observed experimentally by neutron diffraction in 1964 \cite{CribierJacrot64} and later imaged directly by Bitter decoration \cite{EssmannTraeuble67, HessRobinson89}. At low fields, the close-packed triangular flux-line lattice is commonly realized. As the field is increased, however, the strong anisotropy of the Fermi surface \cite{KoganBullock97} and of the superconducting gap \cite{IchiokaHasegawa99} in some unconventional superconductors may lead to pronounced phase transitions to less trivial lattice symmetries as soon as the anisotropic vortex cores are forced into closer proximity \cite{RisemanKealey98, GilardiMesot04, BrownCharalambous04, WhiteBrown08, WhiteHinkov09}.

In many actual materials, nevertheless, pinning of the magnetic flux lines by point-like defects, disorder, or twin domain boundaries prevents the formation of a well-ordered vortex lattice. Instead, a vortex-glass phase characterized only by a short-range positional order is often formed \cite{Fisher89, BlatterFeigelman94, GiamarchiDoussal97, NattermannScheidl00, Huebener01}. The presence of such disorder is not necessarily an issue of sample quality. High-$T_{\rm c}$ superconductors, for example, are obtained by chemical doping of stoichiometric parent compounds with impurity atoms or vacancies, which by themselves may serve as natural pinning centers \cite{AuslaenderLuan08} and enhance pinning even in pristine single crystals.

There are two complementary classes of experimental techniques for visualization of magnetic-flux-line lattices. The first one includes diffraction methods like small-angle neutron scattering (SANS), capable of observing the vortex lattice in the reciprocal space by measuring neutron diffraction patterns that originate from the inhomogeneous field distribution in the whole volume of the sample. The second class includes real-space imaging methods \cite{FasanoMenghini08}, such as Bitter decoration, scanning tunneling microscopy (STM), or magnetic force microscopy (MFM), that can directly probe the local distribution of vortices within a small area on the sample surface.

In the present paper, we apply both SANS and MFM as complementary techniques to study the magnetic-flux-line lattices in slightly overdoped BaFe$_{2-x}$Co$_x$As$_2$ (BFCA) with the doping level $x=0.19$ and the onset superconducting transition temperature $T_{\rm c}=23$\,K. This material belongs to the so-called 122-family of the novel iron arsenide high-temperature superconductors \cite{RotterTegel08, SefatJin08, NiTillman08}. Observations of vortex lattices in BFCA are still limited to a recent STM study on an overdoped ($x=0.2$) sample \cite{YinZech09}, a combination of SANS and Bitter decoration methods applied to a slightly underdoped ($x=0.14$) \cite{EskildsenVinnikov09, EskildsenVinnikov09a}, and an MFM investigation of an underdoped ($x=0.10$) \cite{LuanAuslaender09} single crystal. All these studies revealed a highly disordered flux-line lattice that is suggestive of strong bulk pinning, in agreement with our results presented below. In the following, we will quantify the degree of disorder in such a lattice using statistical analysis of multiple MFM images. We will show that the vortex lattice is characterized by predominantly hexagonal coordination, as expected for a disordered triangular lattice, and will estimate the radial correlation length of such order from our MFM and SANS data in a wide range of the applied magnetic fields.

\vspace{-5pt}\section{Model of the lattice disorder}\vspace{-5pt}\enlargethispage{2pt}

Comparing the correlation length of the vortex lattice extracted from SANS data, measured in the reciprocal space, to the direct-space MFM measurements is rather nontrivial, as different models of disorder may lead to ambiguous definitions of the correlation length. Strictly speaking, no single well-defined parameter can fully characterize the ``degree of disorder'' in vitreous or amorphous lattices with short-range correlations. Therefore, to gain some consistent quantitative understanding from both direct- and reciprocal-space measurements, we start by introducing a continuous Gaussian model of positional disorder applied to flux line displacements perpendicular to the field
direction. This will later prove to be consistent with our experimental results.

Let $\mathbf{r}_{ij}$ be a set of vectors representing a regular periodic two-dimensional (2D) lattice in direct space with a lattice parameter $a$. The disorder is introduced by adding a random vector $\mathbf{w}(\mathbf{r}_{ij})=\{w_{\!x}(\mathbf{r}_{ij}),\,w_{\!y}(\mathbf{r}_{ij})\}$ to every lattice site:\vspace{-0.2em}
\begin{equation}
\mathbf{R}_{ij} = \mathbf{r}_{ij}+\mathbf{w}\kern1pt(\mathbf{r}_{ij}).\vspace{-0.2em}
\end{equation}
The vector components $w_{\!x}(\mathbf{r})$ and $w_{\!y}(\mathbf{r})$ are two independent centered Gaussian stochastic processes of a two-dimensional argument with dispersion given by\vspace{-0.5em}
\begin{multline}\label{Eq:Dispersion}
\forall~\mathbf{r}_1,~\mathbf{r}_2\in\mathbb{R}^2~~\bigl\langle[w_{\!x}(\mathbf{r}_1)-w_{\!x}(\mathbf{r}_2)]^{\,2}\bigr\rangle\\
=\bigl\langle[w_{\!y}(\mathbf{r}_1)-w_{\!y}(\mathbf{r}_2)]^{\,2}\bigr\rangle=\sigma^2|\mathbf{r}_1-\mathbf{r}_2|/a.\vspace{-0.5em}
\end{multline}
Here $\sigma\ll{a}$ is the standard deviation of the distance between nearest-neighbor lattice sites from its mean value $a$. Angle brackets denote mathematical expectation.\smallskip

The \textit{radial distribution function} (RDF) \cite{Ziman79, Ossi03} of such a lattice, $g(r)$, can be well approximated by a sum of Gaussian peaks, such that their widths increase proportionally to $\sqrt{r}$, as follows from the property (\ref{Eq:Dispersion}):
\begin{equation}
g(r)\,= \sum_{n\kern.5pt=1}^\infty \frac{N_n}{\sigma\sqrt{2\piup\,R_n/a}}\,\exp\Biggl[\!-\,\frac{(r-R_n)^2}{2\sigma^2\,R_n/a}\,\Biggr],\quad r>0.
\label{Eq:RDF}
\end{equation}
Here $R_n$ is the radius of the $n^{\rm th}$ coordination shell, and $N_n$ is the number of sites in this shell.

Under these assumptions, on the scale of $L=a^3/(2\sigma\kern-1pt)^2$ the width of the peaks, $2\sigma\!\sqrt{r/a}$, will become comparable with the lattice spacing, and individual peaks in $g(r)$ will merge into a linearly increasing continuum background. The knowledge of $L$ therefore provides an upper boundary for the \textit{range of order}\,---\,a distance beyond which no measurable oscillations in $g(r)$ can be detected \cite{Ziman79}. For $r<L$, the RDF will exhibit oscillatory behavior, with the roughly exponentially decaying amplitude $\sim$\,$\exp(-r/\zeta)$, as we will demonstrate below. Such exponential decay suggests a natural definition for the radial correlation length $\zeta$. We point out that $\zeta$ is several times smaller than the maximal length scale at which detectable correlations exist. More precisely, the following relationship between $\zeta$ and the parameters of our model is derived in the Appendix:
\begin{equation}\label{Eq:zeta_vs_L}
\begin{split}
\zeta&=\frac{\sigma^2}{a}\!\left[\sqrt{\frac{1}{2}+\sqrt{\frac{1}{4}+4\piup^2\,\frac{\sigma^4}{a^4}}}\,-1\right]^{-1}\\
     &=\frac{a^3}{2\piup^2\sigma^2}+\frac{5\sigma^2}{2\kern.5pt a~}+ \sigma\,\mathcal{O}\Bigl[(\sigma/a)^5\Bigr]\approx\frac{2L}{\piup^2}\approx L/5.
\end{split}
\end{equation}
The \textit{reduced radial distribution function} $\widetilde{g}(r)$ can be constructed as \cite{KopitzkiHerzog04} \vspace{-0.5em}
\begin{equation}
\widetilde{g}(r)=g(r)-2\piup r\kern-.5pt g_0,
\end{equation}
where $g_0$ is the average density of sites in the 2D lattice. For a triangular lattice, in particular, $g_0=2/\!\sqrt{3}a^2$. The new function $\widetilde{g}(r)$ has a convenient property \footnote{Though the limit in Eq.\,(\ref{Eq:rRDFlimit}) is exactly zero for the true RDF, the function $\widetilde{g}(r)$, as constructed according to Eq.\,(\ref{Eq:RDF}), will have a small but finite limit of the order of smallness $(\sigma/a)\kern1pt^2$. This finite value arises from the deviation of the actual shape of individual peaks from Gaussian that becomes significant only for large $\sigma$. For our purposes, this deviation can be neglected except in Fig.\,5\,(b), where we explicitly subtract this limit at $r\rightarrow\infty$ to avoid the distortion of the logarithmic plot.}:
\begin{equation}\label{Eq:rRDFlimit}
\lim_{r\rightarrow\infty}\widetilde{g}(r)=0,
\end{equation}
so that its Fourier transform is well defined:
\begin{equation}
S(k)=\int_0^\infty\!\widetilde{g}(r)\sin(kr)\,\mathrm{d}r.
\end{equation}

It can be shown that $S(k)$ is nothing else but the \textit{reduced structure factor}, which can be obtained from the experimentally measured elastic scattering cross-section after form-factor and resolution corrections \cite{KopitzkiHerzog04, Enderby85, Bacon75}. The exponential decay of oscillations in $\widetilde{g}(r)$ ensures a Lorentzian shape of the first diffraction peak in $S(k)$ with the half-width at half-maximum (HWHM) of $\zeta^{-1}$ (for example, see Ref.\,\onlinecite{ZaliznyakLee05}).

\vspace{-5pt}\section{Sample preparation and characterization}\vspace{-5pt}

\begin{figure*}[t]\vspace{-0.5em}
\begin{center}\includegraphics[width=\textwidth]{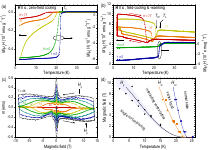}\end{center}
\caption{Magnetization measurements. (a)~dc-susceptibility measured after zero-field cooling in various applied fields up to 7\,T. (b)~dc susceptibility measured during field-cooling and -warming. In both panels, note the different scale for the 1\,mT and 10\,mT data (right axis) and the 1\,T, 3\,T, and 7\,T data (left axis). (c)~Isothermal magnetization loops measured at various temperatures (as indicated beside each curve) in a wide field range up to 14\,T exhibiting the ``fishtail'' effect. Note that the position of the second peak $H_{\rm p}$ is somewhat higher in measurements with increasing field than in the reverse direction. (d)~Magnetic phase diagram summarizing the temperature dependence of the second magnetization peak $H_{\rm p}$, irreversibility field $H_{\rm irr}$, and second critical field $H_{\rm c2}$ extracted from the magnetization measurements. The values of $H_{\rm p}$ measured in increasing and decreasing fields are marked with up- and down-pointing triangles, respectively. Triangles and squares denote values of $H_{\rm irr}$ and $H_{\rm c2}$ measured on two different sample pieces from the same growth batch.}
\label{Fig:Magnetization}
\end{figure*}

The single crystal of BFCA used for the present study has been grown by the self-flux method in a zirconia crucible, using an Al$_2$O$_3$ nucleation center \cite{SunLiu09}. SANS measurements were done on a $\sim$\,0.4\,g as-grown single crystal, whereas a smaller sample from the same growth batch was used for MFM measurements. The doping level of $x=0.19\pm0.01$ was determined by energy-dispersive X-ray analysis (EDX) after averaging over the sample surface. The onset superconducting transition temperature $T_{\rm c}=23$\,K was measured by SQUID magnetometry [Fig.\,\ref{Fig:Magnetization}\,(a)]. These values let us pinpoint our sample on the overdoped side of the phase diagram \cite{NingAhilan09, ChuAnalytis09, WangWu09}. The choice of such doping level is motivated by the non-magnetic ground state of this system, which excludes the possibility of any phase separation \cite{ParkInosov09, InosovLeineweber09, EvtushinskyInosov09, KhasanovEvtushinsky09, GokoAczel09} or microscopic coexistence \cite{MasseeHuang09, DrewNiedermayer09, ChristiansonLumsden09} of the antiferromagnetic and superconducting phases that is observed in underdoped samples and can potentially influence the flux penetration pattern. Since it was proposed that in the underdoped single crystals vortex pinning occurs mainly on the boundaries of intertwinned orthorhombic/antiferromagnetic domains \cite{TanatarKreyssig09, ProzorovTanatar09b, SunLiu09prb}, by choosing an overdoped sample we can exclude this pinning channel from consideration. The absence of static magnetism in our sample has been confirmed by a zero-field low-temperature muon spin relaxation ($\mu$SR) experiment.

Temperature-dependent magnetization measured by SQUID magnetometry after zero-field cooling (ZFC) and during in-field cooling (FC) with subsequent warming are shown in Fig.\,\ref{Fig:Magnetization}\,(a) and (b), respectively. The temperatures at which the FC curves start to deviate from linear behavior correspond to the superconducting transition. The dip at which the hysteresis develops represents the irreversibility temperature $T_{\rm irr}$, below which the flux line lattice gets pinned, whereas at temperatures between $T_{\rm irr}$ and $T_{\rm c}$ depinning of the vortices by thermal fluctuations leads to the melting of the flux line lattice. Combining the ZFC/FC data for various applied fields, the temperature dependencies of the second critical field $H_{\rm c2}$ and of the irreversibility field $H_{\rm irr}$, were extracted.

Isothermal magnetic hysteresis loops were measured at different temperatures in a wide range of magnetic fields up to 14\,T using a \textit{Quantum Design PPMS} extraction sample magnetometer, as shown in panel (c) of the same figure. The non-monotonic behavior with a well-pronounced second peak in magnetization represents the so-called ``fishtail'' effect. It is well known from both conventional and high-temperature superconductors, but its origins are discussed controversially. Most theoretical approaches agree that the temperature-dependent field $H_{\rm p}$, at which the second magnetization peak has its maximum, is related to vortex pinning and corresponds to a crossover between two different regimes of the vortex lattice. Various scenarios have been proposed responsible for the fishtail effect, among them a change in dynamics of the vortex lattice \cite{KrusinElbaum92}, a change in flux creep behavior \cite{AbulafiaShaulov96}, or a change in its elastic properties \cite{BlatterGeshkenbein04}. A fishtail effect of comparable magnitude has already been observed recently in several iron arsenide superconductors \cite{ProzorovNi08, ProzorovTanatar09, ProzorovTanatar09b, YangLuo08, SunLiu09prb}.

The values of all three characteristic fields ($H_{\rm c2}$, $H_{\rm irr}$, and $H_{\rm p}$) define the magnetic phase diagram of the material that is summarized in Fig.\,\ref{Fig:Magnetization}\,(d). The superconducting phase consists of three regions: (i) vortex liquid phase above $H_{\rm irr}$, where magnetic flux lines are not pinned due to strong thermal fluctuations; (ii) region between $H_{\rm p}$ and $H_{\rm irr}$, where intervortex repulsion exceeds the typical pinning forces, presumably leading to an elastically interacting vortex lattice; (iii) low-field region, the nature of which may depend on the actual pinning strength \cite{BlatterGeshkenbein04, SauerzopfWerner00, JirsaKoblischka00}. The measurements presented in the following were performed at low temperatures in a broad range of magnetic fields spanning the region below $H_{\rm p}$ to elucidate the actual structure of the vortex lattice within this phase.

\vspace{-5pt}\section{Small-angle neutron scattering}\vspace{-5pt}

\subsection{Experimental conditions}\vspace{-5pt}

The SANS experiment was carried out using the NG3-SANS diffractometer at NIST. For all measurements presented here, cold neutrons were collimated over a distance of 8.0\,m before the sample. The diffracted neutrons were collected using a two-dimensional multidetector of 128\,$\times$\,128 pixels, each 5.08\,mm\,$\times$\,5.08\,mm, positioned 11.5\,m behind the sample.

The momentum resolution of the instrument ${\scriptstyle\Delta}q\kern.5pt(q)$, expressed in terms of the Gaussian standard deviation, was estimated in the approximation of strong longitudinal disorder, i.e.~large rocking curve width. In this limit, the instrumental contribution to the peak width in the radial direction is given by \cite{CubittForgan92}\vspace{-0.3em}
\begin{equation}\label{Eq:SANS_Resolution}
{\scriptstyle\Delta}q\kern.5pt(q) = \sqrt{\sigma_{\rm\!a}^2\,k^2 + q^2\,({\scriptstyle\Delta}k/k)\kern.5pt^2}
\end{equation}
For our SANS results herein, the standard deviation angular spread of the incoming beam $\sigma_{\rm\!a}$ was measured to be $0.085^\circ$, the mean neutron wave number $k$ was 1.26\,\AA$^{-1}$ with standard deviation spread ${\scriptstyle\Delta}k / k = 6.2$\,\%.

\begin{figure}[b]
\includegraphics[width=\columnwidth]{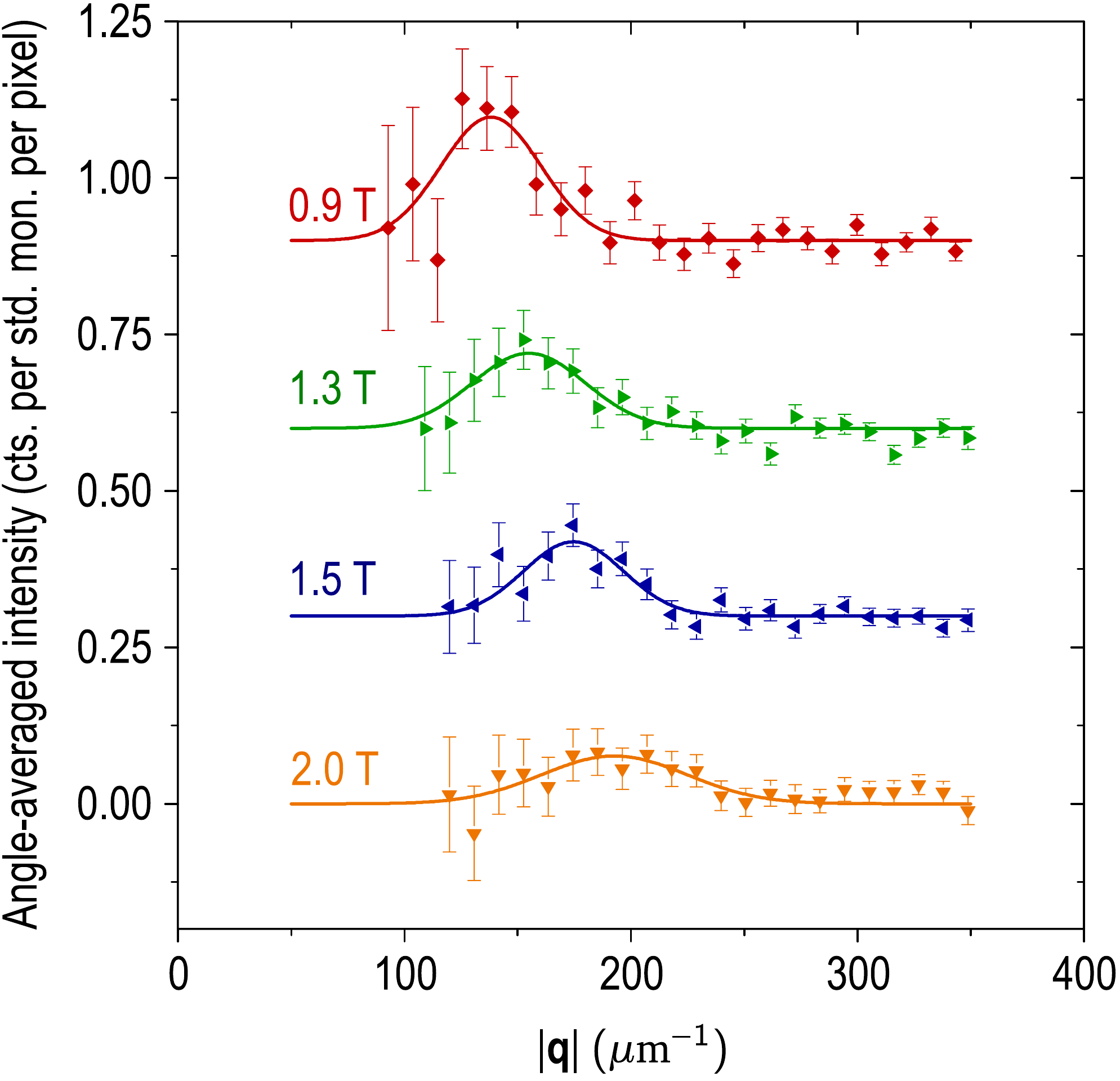}
\caption{Angle-averaged SANS intensity as a function of momentum transfer $|\mathbf{q}|$, measured at different magnetic fields between 0.9\,T and 2.0\,T. For clarity, each curve is offset by an increment of 0.3 units from
the one below it. The vertical scale represents the mean diffracted intensity per pixel, normalized to the standard monitor. The solid lines are Gaussian fits to Eq.\,(\ref{Eq:SANS_Gaussian}).\vspace{-5pt}}
\label{Fig:SANS_G34}
\end{figure}

The single-crystalline sample was loaded into the cold variable-temperature insert of a cryomagnet that provided applied fields of up to 9\,T and a base temperature of 3.5\,K. The sample was oriented with the \textbf{a}-axis vertical, i.e. perpendicular to the applied field, and the \textbf{c}-axis parallel to the field. Our measurements were carried out in the usual SANS experimental geometry, with applied field approximately parallel to the neutron beam. The vortex lattice was prepared in the sample by applying the desired field above $T_{\rm c}$, and subsequent field-cooling to 3.5\,K. In all cases, background measurements were carried out at 26\,K (above $T_{\rm c}$) and subtracted from the field-cooled foreground measurements. An attempt to oscillate the field during cooling to equilibrate the vortex lattice did not result in any notable reduction of lattice disorder. Since such wiggling helps only against a weak pinning background, this fact is consistent with strong vortex pinning in our sample.

\vspace{-5pt}\subsection{SANS results}\vspace{-5pt}

\begin{table*}[t]\normalsize
\begin{center}
\begin{tabular}{c@{~}|@{~~}c@{~~}c@{~~}c@{~~}c@{~~}|@{~~}c@{~~}c@{~~}c}\toprule
$H$    & $\!a_\triangle$~($\mu$m)& $\!a_\square$~($\mu$m) & $4\piup/a_\triangle\!\sqrt{3}$ & $2\piup/a_\square$ & $q_0$ ($\mu$m$^{-1}$)\! & HWHM ($\mu$m$^{-1}$)\! & $\zeta$~($\mu$m) \\
\midrule
0.9\,T & 0.0515 & 0.0480 & 141\,$\mu$m$^{-1}$ & 131\,$\mu$m$^{-1}$ & $138\pm10$ & $26\pm6$  & $0.079\pm0.026$ \\
1.3\,T & 0.0429 & 0.0399 & 169\,$\mu$m$^{-1}$ & 157\,$\mu$m$^{-1}$ & $155\pm12$ & $29\pm8$  & $0.066\pm0.043$ \\
1.5\,T & 0.0399 & 0.0371 & 182\,$\mu$m$^{-1}$ & 169\,$\mu$m$^{-1}$ & $175\pm12$ & $26\pm5$  & $0.068\pm0.027$ \\
2.0\,T & 0.0346 & 0.0322 & 210\,$\mu$m$^{-1}$ & 195\,$\mu$m$^{-1}$ & $193\pm14$ & $38\pm9$  & $0.036\pm0.013$ \\
\bottomrule
\end{tabular}
\end{center}
\caption{Best-fit parameters of the SANS data from Fig.\,\ref{Fig:SANS_G34}. Here $a_\triangle=\sqrt{2\phi_0/H\sqrt{3}}$ and $a_\square=\sqrt{\phi_0/H}$ stand for the parameters of perfect triangular and square vortex lattices, respectively, calculated for every magnetic field, $\phi_0=h/2e$ being the magnetic flux quantum; $4\piup/a_\triangle\!\sqrt{3}$ and $2\piup/a_\square$ are the expected radii of the corresponding diffraction rings that are to be compared with the fitted peak position $q_0$; HWHM of the diffraction peaks are given prior to resolution correction; $\zeta$ is the resolution-corrected radial correlation length.}
\label{Tab:SANSparams}
\end{table*}

Usually in SANS experiments, the sample and cryomagnet are rotated together in order to allow a reciprocal lattice vector of the vortex lattice to satisfy the Bragg condition at the detector. Generally speaking, the Bragg spots in the reciprocal space have a finite size determined both by the instrument resolution and by the structure of the vortex lattice. Therefore, in order to capture all of the diffracted intensity associated with a Bragg spot, the sample is usually rotated through a range of angles about the vertical axis (the \textit{rocking scan}). However, instead of observing distinct Bragg spots, our preliminary measurements on the BFCA sample identified the vortex lattice as strongly disordered; the diffraction signal was weak and took the form of a ring centered on the origin of reciprocal space. This observation is qualitatively identical to the previous studies reported by Eskildsen~\textit{et.\,al.} on a slightly underdoped sample \cite{EskildsenVinnikov09, EskildsenVinnikov09a}. On rotating the sample by $\pm6^{\circ}$ about the vertical axis, within the statistical error there was no variation of the diffracted intensity at any point around the ring. This indicates that the rocking curve width is extremely broad, and hence all subsequent measurements were carried out at a fixed rotation angle. The observation of a ring of scattering indicates the absence of orientational order of the vortex lattice about the field axis. Both of these observations are consistent with strong bulk pinning of the vortices. However, the finite radial width of the diffraction ring indicates a finite planar, or $d$-spacing, order which we can investigate in a more quantitative manner.

In Fig.\,\ref{Fig:SANS_G34}, for a selection of fields we show the radial dependence of the diffracted intensity. At each field, the data were obtained by averaging over $2\piup$ annuli that were centered at the origin of reciprocal space, as defined from the position of the unscattered beam on the detector. For each $|\mathbf{q}|$-bin of a curve, the intensity is obtained from the mean counts per standard monitor per pixel averaged over all the pixels whose location in $\mathbf{q}$ lies within the width of the $|\mathbf{q}|$-bin from the origin. The resulting curves show weak, but clear, peaks in the radial intensity, where the location of the peak in $|\mathbf{q}|$ corresponds to the radius of the diffraction ring in reciprocal space. With increasing field, we clearly see the position of the peak in $|\mathbf{q}|$ increasing, thus confirming that the origin of our diffraction signal is due to a vortex lattice. The solid lines shown in Fig.\,\ref{Fig:SANS_G34} are fits to the following Gaussian function:
\begin{equation}\label{Eq:SANS_Gaussian}
I(q)=A\,\exp\left[-(q-q_0)^2\,\big/\,2\bigl(\zeta^{-2}+{\scriptstyle\Delta}q\text{(}q_0\text{)}^2\bigr)\right].
\end{equation}
The fitting results are summarized in Table\,\ref{Tab:SANSparams}.

Despite our relaxed resolution setup, the observed diffraction peaks are not resolution-limited. Because of these broad widths, it is difficult to draw a confident conclusion about the local symmetry of the vortex lattice solely from the SANS data. Within the uncertainty of our results, the radius of the diffraction ring could be reconciled both with the hexagonal and square coordination (see Table~\ref{Tab:SANSparams}).

In principle, the broad peaks could indicate that the vortex lattice is composed of very small and randomly oriented \textit{anisotropic} domains. Anisotropic vortex structures, for example in the form of distorted hexagonal vortex lattices, are commonly observed in SANS studies of type-II superconductors \cite{RisemanKealey98, GilardiMesot04, BrownCharalambous04, WhiteBrown08, WhiteHinkov09}. They are particularly prominent at higher fields when the vortices start to overlap. In these situations, anisotropies of the superconducting gap \cite{IchiokaHasegawa99}, or of the Fermi surface in combination with non-local effects \cite{KoganBullock97}, can directly cause the deviation from the ideal isotropic and hexagonal vortex lattice. However, any influence of a superconducting gap anisotropy is not expected to be important over the field range explored here \cite{IchiokaHasegawa99}. Furthermore, the anisotropy caused by non-local effects is expected to be suppressed in the presence of disorder, due to a corresponding reduction in the mean free path (reduction in non-locality range). As all the available direct-space measurements, both in previous works \cite{YinZech09, EskildsenVinnikov09, EskildsenVinnikov09a} and this study, point towards isotropic and hexagonal vortex lattice coordinations, we assume the lattice disorder to be the dominant source of diffraction-line broadening in our SANS data. Therefore, the HWHM of the Gaussian lineshapes shown in Fig.\,\ref{Fig:SANS_G34} can be used to measure the radial correlation length $\zeta$ according to Eq.\,(\ref{Eq:SANS_Gaussian}), ignoring possible structural contributions. Subsequent comparison of the extracted values of $\zeta$ with those extracted from direct-space STS measurements performed in somewhat higher fields \cite{YinZech09} will show a reasonable agreement between the two probes, indicating the validity of this approximation.

\vspace{-5pt}\section{Magnetic force microscopy}\vspace{-5pt}

\begin{figure*}[t]
\begin{center}\includegraphics[width=0.85\textwidth]{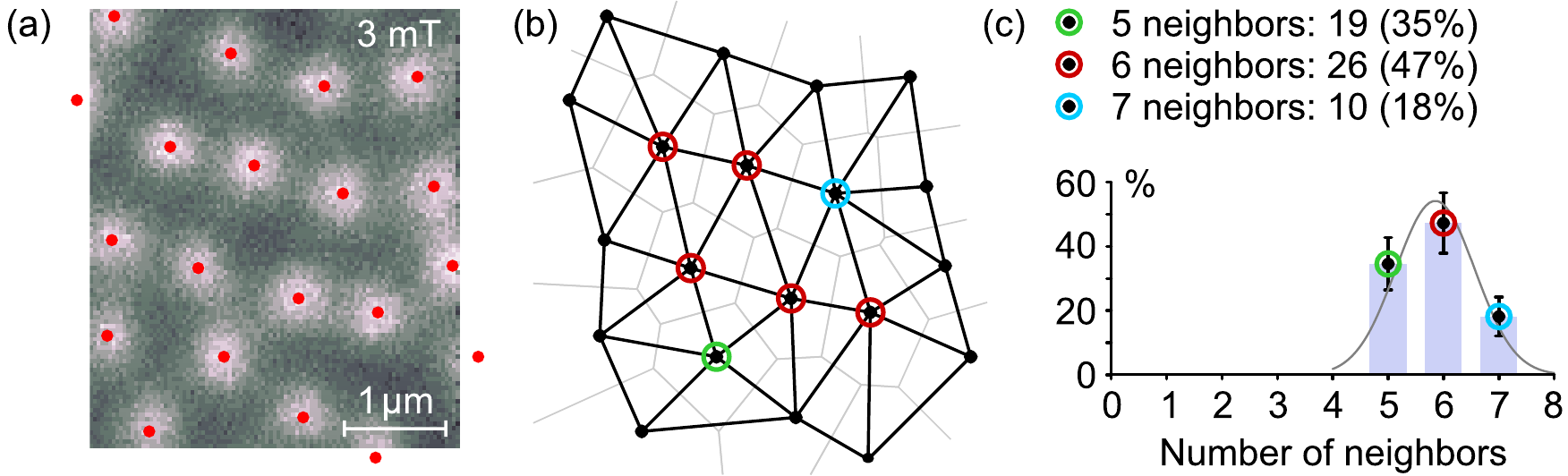}\end{center}
\caption{(a)~One out of eight MFM images measured at 3\,mT with vortex positions (red dots) determined by 2D Gaussian fitting. The lighter colors at the vortex positions correspond to a positive change of the cantilever's resonance frequency. (b)~Delaunay triangulation of the vortex lattice (black lines) and the corresponding Voronoi tessellation (gray lines). The vortices are marked according to the number of nearest neighbors. (c)~Coordination number distribution for all 3\,mT images, showing the predominance of a hexagonal coordination. The thin line is a Gaussian fit of the histogram.}
\label{Fig:Vortices}
\end{figure*}

\subsection{Experimental conditions}\vspace{-5pt}

Low-temperature MFM measurements were conducted using an \textit{Omicron Cryogenic SFM} scanning force microscope \cite{ShapovalNeu07} supplied with a commercial \textit{SSS-MFMR} magnetic tip from \textit{NanoAndMore\,GmbH} with a force constant of $\sim$\,2.8\,N/m and a resonance frequency of 51\,kHz. The measured MFM signal represents a shift of the cantilever's resonance frequency, which is proportional to the normal component of the force gradient acting between the tip and the sample at a given scanning distance above the surface \cite{AlbrechtGruetter91, ProkschDahlberg05}.

A piece of the same single-crystalline sample as the one used for SANS measurements was cleaved \textit{in situ} in high vacuum to get a clean surface. Magnetic fields $H=3$ and 6\,mT were applied along the crystallographic $\mathbf{c}$-axis of the sample in the direction antiparallel to the magnetization of the MFM tip. In such geometry, the interaction between the MFM tip and the superconducting vortices is repulsive and shows up as a positive frequency shift. The sample was then field-cooled down to the base temperature of 8\,K. All measurements were done at a distance of 80\,nm between the tip and the sample. Due to a technical limitation of the microscope, the size of a single MFM image could not be made larger than $4\,\mu{\rm m}\times4\,\mu{\rm m}$ at the base measurement temperature. Therefore, to gather enough statistics, 8 and 4 images were measured at 3 and 6\,mT, respectively, with full temperature cycling before each measurement.

\vspace{-5pt}\subsection{Statistical processing of MFM datasets}\vspace{-5pt}

For each vortex, the 2D vector of its coordinates $\mathbf{v}_i$ was determined by 2D Gaussian fitting, the width being kept equal for all the peaks. This method allows for sub-pixel precision and can successfully locate even those vortices that are partially cut by the image boundary, as shown in Fig.\,\ref{Fig:Vortices}\,(a). Voronoi cells have been calculated for all vortices, i.e. the loci of points that are closer to $\mathbf{v}_i$ than to any other vortex in the set \cite{Dirichlet50, Voronoi07}. The number of sides of each Voronoi cell therefore represents the number of nearest neighbors for the corresponding vortex. Straight lines connecting all nearest-neighbor sites form a Delaunay triangulation \cite{Delaunay34}, as illustrated in Fig.\,\ref{Fig:Vortices}\,(b). Panel (c) of the same figure shows the distribution of vortex coordination numbers for all images measured at 3\,mT. Note that only vortices surrounded by closed Voronoi polygons possess the full set of neighbors and can therefore be counted. The maximum of the distribution is reached for the coordination number of 6, which gives us the first indication of the local lattice symmetry. In the following, we will therefore assume that the vortex phase can be treated as a disordered triangular lattice. More solid evidence supporting this assumption will come from the consideration of the angular distribution function (ADF) in what follows.

\begin{figure*}\vspace{-1em}
\includegraphics[width=\textwidth]{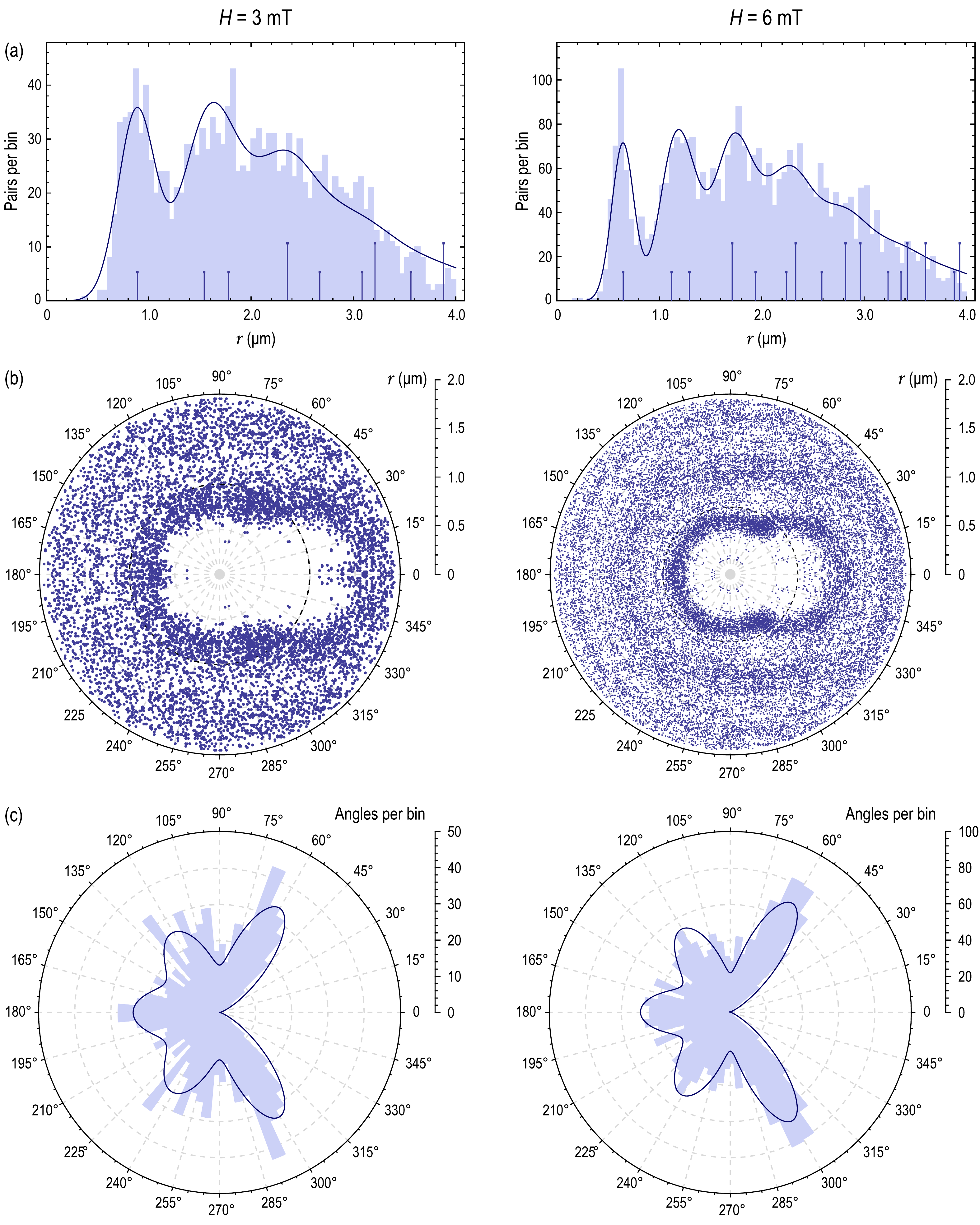}
\caption{Statistical analysis of multiple $4\,\mu{\rm m}\times4\,\mu{\rm m}$ MFM images measured in magnetic fields of 3\,mT (left) and 6\,mT (right). For the two fields, data from 8 and 4 images, respectively, were used. (a) Radial distribution function. Vertical lines mark the positions and relative intensities of individual coordination shell peaks. (b) Triplet distribution function. (c) Angular distribution function.}
\label{Fig:MFMstats}
\end{figure*}

\begin{figure}[b]
\includegraphics[width=\columnwidth]{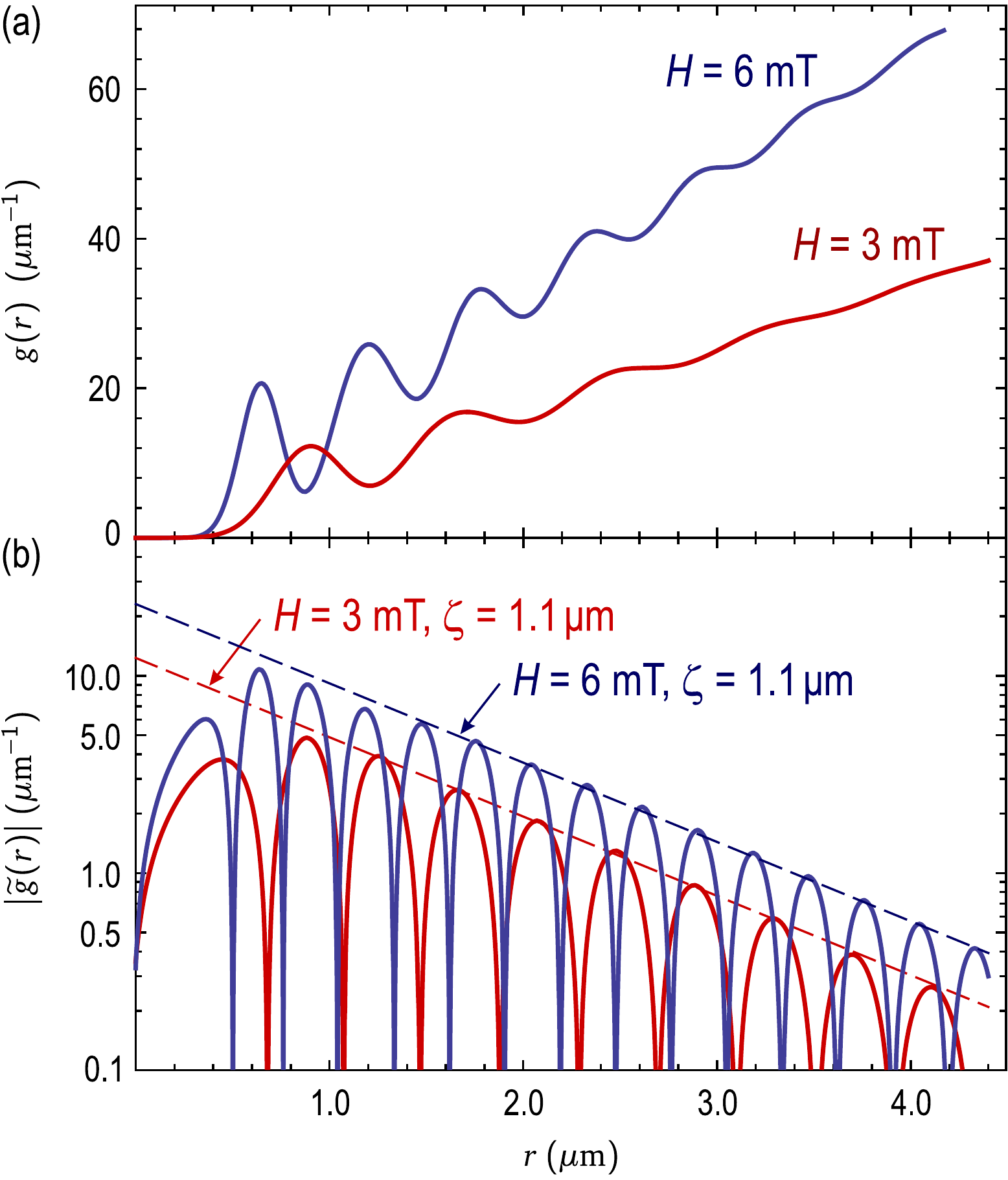}
\caption{Determination of the radial correlation length $\zeta$ from MFM data. (a)~Radial distribution functions $g(r)$ extracted from the fits of the histograms shown in Fig.\,3\,(a). (b)~Absolute value of the reduced RDF
estimated as\vspace{4pt}\\ \mbox{\hfill$\hspace{3em}\widetilde{g}(r)=g(r)-4\piup r\big/\sqrt{3}\,R_1^2-\lim\limits_{r\rightarrow\kern.5pt\infty}\left[g(r)-4\piup r\big/\sqrt{3}\,R_1^2\right]$,\hfill}\vspace{2pt}\\
plotted in logarithmic scale to emphasize the exponential decay of the oscillations. The slope of the exponential envelopes (dashed lines) is a measure of the radial correlation length $\zeta$, which in our case is the same for both fields within the statistical error.\vspace{-2pt}}
\label{Fig:MFMcorrel}
\end{figure}

In the next step, the relative distances $d_{ij}=|\mathbf{v}_i-\mathbf{v}_j|$ were determined for all pairs of vortices $\mathbf{v}_i$ and $\mathbf{v}_j$. The histograms of such distances, which approximate the RDF, are plotted in Fig.\,\ref{Fig:MFMstats}\,(a) for both magnetic field values. Solid lines show fits to\vspace{-0.5em}
\begin{equation}
\frac{\frac{1}{\raisebox{1.5pt}{${\scriptstyle 2}$}}\,N g(r)\,\delta\!r}{1+\exp[(r-R_{\rm max})/W_{\rm max}]},
\end{equation}
where $N$ is the total number of vortices in the dataset, $g(r)$ is the RDF of a triangular lattice given by Eq.\,(\ref{Eq:RDF}), $\delta\!r$ is the histogram's bin size, and the factor of 1/2 is added to avoid double counting of equivalent vortex pairs ($d_{ij}=d_{ji}$). The denominator is an empirical factor introduced to compensate for the distribution function cut-off at large $r$ due to a finite image size. The best-fit parameters of the model are given in Table~\ref{Tab:MFMparams}. The pristine RDFs deprived of the experimental correction factors are also plotted in Fig.\,\ref{Fig:MFMcorrel}\,(a).

\begin{table*}[t]
\begin{center}\normalsize
\begin{tabular}{c|c|c|@{~}c@{~~}c@{~~}c@{~~}c|c@{}c|c}\toprule
$H$     & $N$     &$\,a_\triangle=\sqrt{2\phi_0/H\sqrt{3}}\,$ & $R_1$ & $\sigma$ & $R_{\rm max}$ & $W_{\rm max}$ & $\zeta$ & $L=R_1^3/(2\sigma\kern-1pt)^2$ & $\sigma\!_\phi$~(deg)\\
\midrule
3.0\,mT & 166     & 0.893\,$\mu$m               & 0.90          & 0.19              & 1.72                   & 0.81                & $1.1\pm0.2$        & $5.1\pm0.4$ & $15.1\pm1.0$\\
6.0\,mT & 163     & 0.631\,$\mu$m               & 0.65          & 0.11              & 1.82                   & 0.73                & $1.1\pm0.2$        & $5.7\pm0.4$ & $14.2\pm1.0$\\
\bottomrule
\end{tabular}
\end{center}
\caption{Best-fit parameters of the RDF and ADF extracted from MFM data and the derived values of the radial correlation length $\zeta$ and the range of order $L$. All values are given in $\mu$m, unless otherwise specified. The lattice constant $a_\triangle$, shown in the second column, is the expected value calculated for a perfect triangular lattice. It is to be compared with the measured radius of the first coordination shell $R_1$. The statistical errors of $\zeta$, $L$, and $\sigma\!_\phi$ were estimated as the standard deviation of values obtained for different binning of the histograms.}
\label{Tab:MFMparams}
\end{table*}

The most pronounced peak appearing at $r=R_1$ corresponds to the first coordination shell of the vortex lattice and can be clearly seen in both histograms. A good agreement is found between the radius of the first coordination shell and the expected lattice constant $a=\sqrt{2\phi_0/H\sqrt{3}}$ calculated for a perfect triangular vortex arrangement (see Table~\ref{Tab:MFMparams}). Here $\phi_0=h/2e=2.07\times10^{-15}$\,T\,m$^2$ is the magnetic flux quantum. The ratio $L=R_1^3/(2\sigma\kern-1pt)^2$ that provides an upper boundary for the range of order in the lattice is the same for both fields within the statistical error. The direct estimation of the correlation length from the RDF, illustrated in Fig.\,\ref{Fig:MFMcorrel}\,(b), also results in the value of $\zeta=1.1\pm0.2\,\mu$m for both fields. This value was obtained by extracting the reduced RDF from the experimental fits, plotting its absolute value in logarithmic coordinates and fitting the positions of its numerous maxima with an exponential-decay function $\exp\kern1pt(A-r/\zeta)$, as shown by the dashed lines in the figure.

Note that the resulting value of $\zeta$ is nearly a factor of 5 smaller than the range of order $L$, in agreement with Eq.\,(\ref{Eq:zeta_vs_L}). Therefore, distinct correlation peaks can be still seen in the RDF histograms [Fig.\,\ref{Fig:MFMstats}\,(a)] even at distances of about 2 to 3 correlation lengths, depending on statistics. For example, the second peak appearing around 1.6\,$\mu$m in the 3\,mT data and around 1.2\,$\mu$m in the 6\,mT data corresponds to the unresolved second and third coordination shells of a triangular lattice, as indicated by the corresponding coordination shell radii shown by vertical lines. The height of these lines is proportional to the number of sites in each shell.

Next, we study the angular distribution of the vortex positions. For this, all possible unordered vortex triples $\{\mathbf{v}_i,\,\mathbf{v}_j,\,\mathbf{v}_k\}$ are considered in every MFM image, such that $\mathbf{v}_j$ lies within the first coordination shell from $\mathbf{v}_i$. The threshold for the distance between vortices to achieve this limitation was chosen to include most of the first-shell peak in RDF (i.e. somewhat above the maximum of the peak). The optimal threshold value for our case was found by trial and error to be $\sim$\,1.0\,$\mu$m at 3\,mT and $\sim$\,0.75\,$\mu$m at 6\,mT. For all such triples, we plot a point in polar coordinates [Fig.\,\ref{Fig:MFMstats}\,(b)] at the distance $|\mathbf{v}_k|$ from the origin and at an angle $\phi$ formed by the vectors $\mathbf{v}_j-\mathbf{v}_i$ and $\mathbf{v}_k-\mathbf{v}_i$. Each point is plotted twice: for $\phi$ and $-\phi$. The dashed circle shows the threshold of the first coordination shell. In such angle-resolved representation, the density of points represents the triple-distribution function (TDF), so that diffuse clusters of points forming around regular lattice sites can be seen. The above-discussed RDF $g(r)$ can essentially be obtained by angle-integration of the TDF.

Now it is straightforward to obtain the ADF of the first coordination shell by plotting a histogram of those angles $\phi$ that correspond to points lying within the threshold radius from the origin (i.e. within the dashed circle). Such histograms are shown in Fig.\,\ref{Fig:MFMstats}\,(c). They immediately reveal a hexagonal coordination of the lattice, as indicated by the pronounced peak located at $\pm60^\circ$ followed by a dip at $\pm90^\circ$. This lets us conclude that the vortex glass phase can be considered as a highly disordered triangular lattice. By analogy with Eq.\,(\ref{Eq:RDF}), the function used to fit the ADF histogram is a sum of Gaussian peaks of the form\vspace{-0.3em}
\begin{equation}
\alpha(\phi)\,= \frac{6\,A\,N\delta\!\phi}{\sigma\!_\phi\sqrt{2\piup}}\,\sum_{n\kern.5pt=1}^5\,\frac{1}{\sqrt{2\sin\kern1pt(\kern-1pt\piup n/6)}}\,\mathrm{e}^{\textstyle-\frac{(\phi-\piup n/3)^2}{4\kern.5pt\sigma_{\!\!\phi}^2\sin\kern1pt(\kern-1pt\piup n/6)}}.\vspace{-0.2em}
\label{Eq:ADF}
\end{equation}
In this model, $\sigma\!_\phi\approx\sqrt{2}\,\sigma\kern-.5pt/\kern-.5pt{a}$ is the standard deviation of the first peak located at $\pm60^\circ\!$. The dispersion of the two other peaks at $\pm120^\circ$ and $180^\circ$ is assumed to increase as $2\sigma_{\!\!\phi}^2\sin(\phi/2)$, i.e.~proportionally to the distance between endpoints of two unit vectors separated by the angle $\phi$, in accordance with Eq.~(\ref{Eq:Dispersion}). The correction coefficient $A<1$ is introduced to compensate for the finite image size effects, and $\delta\!\phi$ is the histogram's bin size.

The only physically meaningful fitting parameter in this model is $\sigma_{\!\phi}$, being just another measure for the degree of randomness in the lattice. The convenience of this parameter is that it is dimensionless, unlike $\sigma$, $\zeta$, or $L$, so it does not require comparison to other characteristic length scales in the system. Again, the values of $\sigma_{\!\phi}$ extracted from the ADF histograms (see Table~\ref{Tab:MFMparams}) turn out to be nearly the same for 3\,mT and 6\,mT within the statistical error.

Here we note in passing that the presented method of ADF analysis is very efficient for discerning the local symmetry of a short-range order from direct-space measurements. Because the position of the first peak in such a plot differs by as much as 30$^\circ$ for the triangular and square lattices, which in our case corresponds to $\sim\!2\sigma_{\!\phi}$, it provides the optimal way to distinguish between the two symmetries. For comparison, the ratio of the lattice constants for the triangular and square lattice symmetries is just $a_\triangle/a_\square=\sqrt{2\kern1pt/\kern-.5pt\sqrt{3}}\approx1.075$. Detecting the difference of $7.5$\% in the radius of the first coordination shell $R_1$ would be impossible under the same conditions.

\vspace{-8pt}\section{Summary and discussion}\vspace{-5pt}\enlargethispage{1pt}

\begin{figure}\vspace{-0.5em}
\begin{center}\includegraphics[width=\columnwidth]{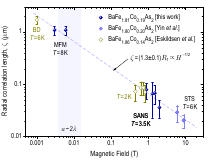}\end{center}\vspace{-1.2em}
\caption{Low-temperature radial correlation length $\zeta$ of the vortex glass phase summarized as a function of the magnetic field $H$. Values extracted from the high-field STS data of Y.\,Yin~\textit{et~al.} \cite{YinZech09} and from the SANS and Bitter decoration (BD) data of M.\,R.\,Eskildsen~\textit{et~al.} \cite{EskildsenVinnikov09, EskildsenVinnikov09a} are added for comparison. Within the shaded region the vortex-lattice constant $a$ becomes comparable with the penetration depth $2\lambda$. The width of this region reflects an uncertainty of the reported penetration depth values \cite{GordonNi09, ProzorovTanatar09a, WilliamsAczel09}. A fit to the power law $\zeta\propto H^{-1/2}$ is shown by the dashed line.\vspace{-0.3em}}
\label{Fig:CorrLen}
\end{figure}

The radial correlation length $\zeta$ of the vortex glass phase is plotted in Fig.\,\ref{Fig:CorrLen} as a function of the applied magnetic field ranging from 1\,mT to 9\,T. In addition to the values extracted from our MFM and SANS data, the figure also includes correlation lengths derived from the high-field STS data of Y.\,Yin~\textit{et~al.} \cite{YinZech09} as well as the SANS and low-field Bitter decoration data of M.\,R.\,Eskildsen~\textit{et~al.} \cite{EskildsenVinnikov09, EskildsenVinnikov09a} STS and Bitter decoration data were analyzed by the same above-described algorithm that we used for statistical MFM image processing. All presented data were measured on samples with doping levels similar to ours and reveal a general power-law trend in the field-dependent correlation length that persists over nearly four decades of the applied magnetic field: \begin{equation}\zeta\approx(1.3\pm0.1)\,R_1=(1.3\pm0.1)\,a\propto H^{-1/2}\end{equation}
Such universality implies that the disorder of the flux line lattice is sample-independent and can possibly be considered an intrinsic property of the as-grown BFCA superconductors close to the optimal doping. The good agreement between correlation length values measured by direct- and reciprocal-space probes confirms the consistency of the two methods.

Our observations are difficult to reconcile with the collective pinning model \cite{LarkinOvchinnikov79}, according to which an increase of the correlated volume with the applied magnetic field is expected \cite{KrusinElbaum92, Tinkham04} that would lead to deviations from the simple power law scaling of $\zeta$ at high fields. Since the correlation radius $R_{\rm c}$ (transversal size of the vortex bundle) predicted in the collective pinning model is inversely proportional to the mean square of the pinning force \cite{Tinkham04}, in the limit of strong pinning it should become smaller than the vortex-lattice constant, leading to the breakdown of the collective model. In this limiting case of single-vortex pinning, vortices no longer form bundles, but behave individually. The ratio $\zeta/a$ then remains at a saturated value of the order of unity, in agreement Fig.\,\ref{Fig:CorrLen}. Thus, the observed field dependence of the radial correlation length can be understood in the limit of strong single-vortex pinning \cite{SauerzopfWerner00, BlatterGeshkenbein04}, meaning that the typical pinning forces dominate over the intervortex repulsion \cite{BlatterGeshkenbein04} in the studied field range. Persistence of such behavior up to the fields comparable with $H_{\rm p}$ at low temperatures suggests that the second maximum in magnetization possibly represents a crossover from the single-vortex pinning regime to an elastically interacting vortex lattice \cite{Kramer73, JirsaKoblischka00}.

For comparison, in the low-$T_{\rm c}$ superconductor SnMo$_6$S$_8$, where the vortex glass phase has been imaged recently in direct space by STM \cite{PetrovicFasano09}, the temperature-dependent crossover from the single to collective pinning regimes was observed at much lower magnetic fields below $H_{\rm x}(T\rightarrow0)=\,2.5$\,mT, well separated from $H_{\rm p}$. In addition, the positional correlation functions in this material exhibited a slower power-law decay, expected for a weakly disordered Bragg glass \cite{KimYao99, FasanoMenghini08}, in contrast to the exponential decay that we observe in BFCA.

The origin of such anomalously strong pinning in doped iron arsenide superconductors and the ways to influence it artificially remain to be investigated. In a recent work, Prozorov \textit{et al.} \cite{ProzorovTanatar09b} argue that in an underdoped BFCA the pinning occurs mainly on the boundaries of intertwinned orthorhombic/antiferromagnetic domains \cite{TanatarKreyssig09}. Our work, however, indicates that pinning still remains anomalously strong even in the overdoped single crystals, where static magnetism is fully suppressed. On the one hand, this could be an evidence in favor of a purely chemical origin of the pinning landscape. On the other hand, evidence for some kind of magnetic order induced by weak magnetic fields in overdoped iron arsenides has been reported from $\mu$SR experiments \cite{KhasanovMaisuradze09}. Therefore the influence of magnetism on the pinning potential cannot be fully excluded.

\vspace{-10pt}\section*{Acknowledgements}\vspace{-5pt}

The authors are grateful to A.~Leineweber, E.\,H.~Brandt, O.\,M.\,Auslaender, H.~Suderow, Y.~Li, and S.\,V.~Inosov for helpful discussions and to B.\,Baum for sample preparation. This work was partially funded in the framework of the DFG-Forschergruppe FOR\,538. SANS experiments were done with financial assistance from the EPSRC UK and with the use of facilities supported in part by the National Science Foundation under Agreement No.\,DMR-0454672. MFM measurements were supported by the European Community under the 6$^\text{th}$ Framework Programme Contract No. 516858: HIPERCHEM.

\bibliographystyle{phpf}\bibliography{VortexGlass}

\vspace{1em}\section*{Appendix}\vspace{-5pt}

To derive the relationship (\ref{Eq:zeta_vs_L}) between the parameters of our disorder model and the radial correlation length $\zeta$, let us consider the simplified case of a one-dimensional periodic lattice with spacing $a$, for which the RDF (\ref{Eq:RDF}) will take the form\vspace{-0.5em}
\begin{equation*}
g(r)\,= \frac{1}{\sigma}\sqrt{\frac{2}{\piup}}\,\sum_{n\kern.5pt=1}^\infty \frac{1}{\sqrt{n}}\,\mathrm{e}^{\textstyle-\frac{(r-na)^2}{2\kern.5pt\sigma^2n}},\quad r>0.
\end{equation*}
Let us now calculate the Fourier transform of the function $h(r)=g(r)\exp(-r\kern-.5pta\kern-.5pt/\kern-.5pt\sigma^2)$:
\begin{equation*}
\hat{h}(k)\,= \sqrt{\frac{2}{\piup}}\,\sum_{n\kern.5pt=1}^\infty\,\mathrm{e}^{\textstyle\frac{-n(a^2+k^2\sigma^4)}{2\sigma^2}}
= \frac{\sqrt{2/\kern-1pt\piup}}{\mathrm{e}^{\textstyle\frac{a^2+k^2\sigma^4}{2\sigma^2}}\!-1}.
\end{equation*}
We can view $\hat{h}(k)$ as a meromorphic function of complex variable $k$. This function has poles at points $k_m=k_m'+\mathrm{i}\kern.5ptk_m''\!\in\mathbb{C}$ which satisfy
$$
a^2+k_m^2\sigma^4 = 4\piup\mathrm{i}\kern.5ptm\sigma^2,~~m\in\mathbb{Z}.
$$
Solving this equation and considering only poles with positive imaginary parts, one gets
\begin{equation*}
\begin{split}
k_m'=\,\frac{ma}{\sqrt{2}\,|m|\sigma^2}&\sqrt{-1+\sqrt{1+\frac{16\kern1ptm^2\piup^2\sigma^4}{a^4}}}~~{\rm and}\\
k_m''=\,\frac{a}{\sqrt{2}\sigma^2}&\sqrt{1+\sqrt{1+\frac{16\kern1ptm^2\piup^2\sigma^4}{a^4}}}
\end{split}
\end{equation*}
Now $h(k)$ can be presented as an inverse Fourier transform of $\hat{h}(k)$ and expanded into a series using the Cauchy residue theorem:\vspace{-0.8em}
\begin{multline*}
\hspace{-1em}h(r)=\frac{1}{\sqrt{2\piup}}\int_{-\infty}^{\infty}\!\hat{h}(k)\,\mathrm{e}^{\mathrm{i}kr}\mathrm{d}k=\frac{2\piup\mathrm{i}}{\sqrt{2\piup}}\sum_{m=-\infty}^\infty\underset{\phantom{_m}k=k_m}{\operatorname{Res}}\!\left[\,\hat{h}(k)\,\mathrm{e}^{\mathrm{i}kr}\right]\\
=\frac{2}{a}\,\mathrm{e}^{-ra\kern-.5pt/\kern-.5pt\sigma^2}+4\sum_{m=1}^\infty\frac{\mathrm{e}^{-k_m''r}}{\sigma^2|k_m|^2}\left[k_m''\cos(k_m'r)-k_m'\sin(k_m'r)\right].
\end{multline*}
Since $k_m''$ monotonically increases with $m$, in the limit $r\rightarrow\infty$ the term corresponding to $m=1$ will be dominant, therefore asymptotically\vspace{-1ex}
\begin{multline*}
g(r)=\frac{2}{a}+\frac{4\,\mathrm{e}^{r\kern-.5pta\kern-.5pt/\kern-.5pt\sigma^2-k_1''r}}{\sigma^2|k_1|^2}\left[k_1''\cos(k_1'r)-k_1'\sin(k_1'r)\right]\\
+\frac{a}{\sigma^2}\,\mathcal{O}(\mathrm{e}^{r\kern-.5pta\kern-.5pt/\kern-.5pt\sigma^2-k_2''r}).
\end{multline*}
The exponent in the second term gives us the sought correlation length:\vspace{-1ex}
$$
\zeta=\frac{1}{k_1''-a\kern-.5pt/\kern-.5pt\sigma^2}=\frac{\sigma^2}{a}\!\left[\sqrt{\frac{1}{2}+\sqrt{\frac{1}{4}+4\piup^2\,\frac{\sigma^4}{a^4}}}\,-1\right]^{-1}.
$$
Expanding this expression in powers of $\sigma/a$, one gets the following approximate formula:
$$
\zeta=\frac{a^3}{2\piup^2\sigma^2}+\frac{5\sigma^2}{2\kern.5pt a~}+ \sigma\,\mathcal{O}\Bigl[(\sigma/a)^5\Bigr]\approx\frac{2L}{\piup^2}\approx L/5.
$$\vfill
\end{document}